\documentclass[conference]{IEEEtran}
\usepackage{balance}
\usepackage{float}
\usepackage{standard}
\usepackage{etex}

\usepackage{rotate}
\usepackage{algorithmic}
\usepackage{algorithm}
\usepackage{float}
\usepackage{tikz}
\usepackage{refcount}

\usepackage{pgf,tikz}
\usepackage{pgfplots}
\usepackage{pgfplotstable}
\usepackage{bbm}
\usepackage{adjustbox}
\usetikzlibrary{calc}
\usepackage{relsize}
\usepackage{ifthen}

\usetikzlibrary{calc,fit,arrows,plotmarks,intersections,patterns,shapes,decorations,positioning, backgrounds, matrix}

\usepackage[numbers,sort&compress]{natbib}
\usepackage[ngerman, english]{babel}

\usetikzlibrary{arrows.meta}
\edef\wdArrowLength{2}
\tikzset{>={Latex[width=1.5mm,length=\wdArrowLength mm]}}
\usepackage{graphicx}

\usepackage{cancel}
\usepackage{mathtools}
\usepackage{shadethm}
\usepackage{empheq}
\usepackage{skull}
\usepgfplotslibrary{units}
\usepackage{calc}
\usepackage{nicefrac}
\usepackage{fancybox}
\usepackage{psfrag}
\usepackage{dsfont}

\title{Sacrificing CSI for a Greater Good:\\ RIS-enabled Opportunistic Rate Splitting
\thanks{This work was funded in part by the Federal Ministry of Education and Research (BMBF) of the Federal Republic of Germany (F\"orderkennzeichen 16KIS1235, MetaSEC). K. Weinberger and A. Sezgin are with the Ruhr-Universit\"at Bochum, Germany Email:\{kevin.weinberger, aydin.sezgin\}@rub.de.}} % Title of the Thesis
\author{Kevin Weinberger, Aydin Sezgin}
\date{\today}

\usepackage{graphicx}
\usepackage{placeins}
\usepackage{booktabs}
\usepackage{marvosym}
\usepackage{multirow}

\usepackage{latexsym, amsmath, amssymb, amsfonts, upgreek}
\usepackage[nolist]{acronym}

\usepackage{tikz}
\usetikzlibrary{calc,arrows,positioning,decorations,shadows,shapes,fadings,matrix}
\tikzset{>=latex'}
\tikzset{semithick}

\makeatletter
\providecommand{\IfElsePackageLoaded}[3]{\@ifpackageloaded{#1}{#2}{#3}}
\makeatother

\usepackage{subfigure}
\IfElsePackageLoaded{subfig}
	{\usepackage[subfigure]{tocloft}}{	
	\IfElsePackageLoaded{subfigure}
		{\usepackage[subfigure]{tocloft}}
		{\usepackage{tocloft}}
	}

\makeatletter

\def\tikz@delimiter#1#2#3#4#5#6#7#8{%
	\bgroup
		\pgfextra{\let\tikz@save@last@fig@name=\tikz@last@fig@name}%
		node[outer sep=0pt,inner sep=0pt,draw=none,fill=none,anchor=#1,at=(\tikz@last@fig@name.#2),#3]
		{%
			{\nullfont\pgf@process{\pgfpointdiff{\pgfpointanchor{\tikz@last@fig@name}{#4}}{\pgfpointanchor{\tikz@last@fig@name}{#5}}}}%
			\delimitershortfall\z@% as suggested by morbusg (maximum space not covered by a delimiter = 0)
			\resizebox*{!}{#8}{$\left#6\vcenter{\hrule height .5#8 depth .5#8 width0pt}\right#7$}%
		}
		\pgfextra{\global\let\tikz@last@fig@name=\tikz@save@last@fig@name}%
	\egroup%
}

\tikzset{hexagon/.code={
	\draw (0,2) -- (-4,0) -- (0,-2) -- (4,0) -- (0,2);
}}

\tikzset{phone/.code={
   \node [rectangle,rounded corners=1.5pt,draw,minimum height=0.6cm, minimum width=0.35cm] at (0,0){};
   \node [rectangle,rounded corners=1.5pt,draw,minimum height=0.5cm, minimum width=0.3cm] at (0,0){};
}}

\makeatletter
\def\cantox@vector#1#2#3#4#5#6#7#8{%
  \dimen@.5\p@
  \setbox\z@\vbox{\boxmaxdepth.5\p@
   \hbox{\kern-1.2\p@\kern#1\dimen@$#7{#8}\m@th$}}%
  \ifx\canto@fil\hidewidth  \wd\z@\z@ \else \kern-#6\unitlength \fi
  \ooalign{%
    \canto@fil$\m@th \CancelColor
    \vcenter{\hbox{\dimen@#6\unitlength \kern\dimen@
      \multiply\dimen@#4\divide\dimen@#3 \vrule\@depth\dimen@\@width\z@
      \vector(#3,-#4){#5}%
    }}_{\raise-#2\dimen@\copy\z@\kern-\scriptspace}$%
    \canto@fil \cr
    \hfil \box\@tempboxa \kern\wd\z@ \hfil \cr}}
\def\bcancelto#1#2{\let\canto@vector\cantox@vector\cancelto{#1}{#2}}
\makeatother

\newcommand{\ifthen}[2]{\ifthenelse{#1}{#2}{}}

\newcommand{\vectw}{\vect{\omega}}

\newcommand{\listequationsname}{List of Formulas}
\newlistof{myequations}{equ}{\listequationsname}

\definecolor{myblue1}{rgb}{0,0,255}
\definecolor{myblue2}{rgb}{65,105,225}
\definecolor{myblue3}{rgb}{70,130,180}
\definecolor{myblue4}{rgb}{176,196,222}

\newcommand{\mytilde}{{\raise.17ex\hbox{$\scriptstyle\mathtt{\sim}$}}}
\newcommand{\naive}{}
\def\naive/{na\"{\i}ve}

\newcommand{\executeiffilenewer}[3]{%
\ifnum\pdfstrcmp{\pdffilemoddate{#1}}%
{\pdffilemoddate{#2}}>0%
{\immediate\write18{#3}}\fi%
}
\pgfkeys{/pgf/fpu}
\pgfmathparse{16383+1}

\pgfkeys{/pgf/fpu=false}

\IEEEoverridecommandlockouts

\begin{document}
\maketitle
\begin{abstract}
In reconfigurable intelligent surface (RIS)-assisted systems, the optimization of the phase shifts requires separate acquisition of the channel state information (CSI) for the direct and RIS-assisted channels, posing significant design challenges. In this paper, a novel scheme is proposed, which considers practical limitations like pilot overhead and channel estimation (CE) errors to increase the net performance. More specifically, at the cost of unpredictable interference, a portion of the CSI for the RIS-assisted channels is sacrificed in order to reduce the CE time. By alternating the CSI between coherence blocks and employing rate splitting, it becomes possible to mitigate the interference, thereby compensating the adverse effect of the sacrificed CSI. Numerical simulations validate that the proposed scheme exhibits better performance in terms of achievable net rate, resulting in gains of up to 160\% compared non-orthogonal multiple access (NOMA), when CE time and CE errors are considered.
\end{abstract} 
\thispagestyle{empty}
\pagestyle{empty}

\section{Introduction}
The \ac{RIS} is a surface that is composed of a large number of passive reflect elements, which enable the ability to shape beams from incident signals \cite{intro1}. This is achieved by controlling the phase shifts of each element individually, enhancing the signals constructively at intended users and destructively at unintended users. However, due to the passive nature of the \ac{RIS}, new challenges arise regarding the separate acquisition of the \ac{CSI} of direct and reflected links, which is a necessity for beneficial adjustment of the \ac{RIS}. State-of-the-art \ac{CE} methods, which support separate \ac{CSI} acquisition, require a prohibitively high number of pilot symbols to do so, due to the proportionality to the number of deployed elements \citep{OnOFF,MISO_CE,Hanzo}.

To overcome this limitation, we develop a transmission scheme, which essentially cuts the proportionality to the number of reflect elements in half, thereby reducing the time required for \ac{CE} substantially. Enabled by the controllability of the \ac{RIS}, the known channels are used for enhancing the channel strength through proper selection of the phase shifts at the cost of unpredictable interference caused by the unknown channels. However, alternating the \ac{CSI} knowledge between coherence blocks facilitates the use of \ac{RS} \cite{aydinWSA,RS_alaa}, which is able to negate the harmful impact of the interference, while interacting synergistically with the \ac{RIS} \cite{weinberger2021synergistic,synergConf}. In fact, we show that the proposed scheme reaps the benefits of both, the increased \ac{DL} time as well as \ac{RS}.\vspace{-0.2cm} 
\section{System Model}\label{ch:Sysmod}
In this work, we consider \ac{RIS}-assisted 2-user communications in a single cell, where the \ac{RIS} is deployed to improve the \ac{DL} communications between a multi-antenna \ac{BS} and a set of $K=2$ single-antenna users $\mathcal{K}\in\{1,2\}$. The number of transmit antennas at the \ac{BS} and the reflect elements of the \ac{RIS} are denoted with $L$ and $N$, respectively. We assume that the \ac{RIS} is equipped with a smart controller enabling individual real-time adjustments of the \ac{RC} at each element \cite{RIS_meta_realtime}. Furthermore, the quasi-static flat-fading model is adopted for all links.
The direct \ac{DL} channels to user $k$ in \ac{CB} $t$ are denoted as $\vect{h}_{t,k} \in \mathbb{C}^{{L}\times 1}$. Similarly, the channels spanning from \ac{BS} to \ac{RIS} and \ac{RIS} to user $k$ in block $t$ are denoted as $\mat{U}_{t}\in \mathbb{C}^{{L}\times N}$ and $\vect{q}_{t,k} \in \mathbb{C}^{{N}\times 1}$, respectively. Furthermore, by denoting the phase shifts at the \ac{RIS} in $t$ as $\vect{\theta}_t\in \mathbb{C}^{N\times 1}$
and the transmit signal vector as $\vect{x}_t \in \mathbb{C}^{L \times 1}$, the received signal at user $k$ can be written as \vspace{-0.15cm}
\begin{align}\label{eq:recSig}
   r_{t,k} &= (\vect{h}_{t,k} + \underbrace{\mat{U}_{t}\,\text{diag}({\vect{q}_{t,k}})}_{\mat{H}_{t,k}}\,\vect{\theta}_t)^{\mathsf{H}}\vect{x}_t + v_{t,k},\\[-20pt]\nonumber
\end{align}
where $\mat{H}_{t,k}$ denotes the cascaded \ac{BS}-\ac{RIS}-user channels to user $k$ and $v_{t,k} \sim \mathcal{CN}(0,\sigma_{v}^2)$ is the receiver noise.
%\subsection{Channel Estimation Protocol}
In this work, we aim to optimize the \ac{RC} at the \ac{RIS}, which necessitates the \ac{CSI} acquisition of each direct and reflected channel within the system. To this end, we utilize the generalized \ac{DFT}-based \ac{CE} method introduced in \cite{Hanzo}, where the quality of the estimates $\hat{\vect{h}}_{t,k}$ and $\hat{\mat{H}}_{t,k}$ of $\vect{h}_{t,k}$ and $\mat{H}_{t,k}$, respectively, is dependent on the noise power at the \ac{BS} $\sigma_z^2$ and the \ac{UL} pilot power $P_{\mathsf{UL}}$. In order to estimate all channels, this method requires at least $\tau_\mathsf{all}=(N+1)K$ pilot symbols. This can potentially limit the application of this \ac{CE} method for scenarios, where a large number of reflect elements is deployed.\vspace{-0.05cm}

\section{Proposed Opportunistic Rate Splitting (ORS)}
In this section, we derive the ORS scheme. By utilizing this scheme, we reduce the duration of the \ac{CE} time to $\tau_\mathsf{half}= (\frac{N}{2} + 1)K$ by neglecting the \ac{RIS}-assisted channels of one user in each \ac{CB} in an alternating fashion. This not only increases the available time for the \ac{DL} data transmission per \ac{CB}, but also imparts new properties to the system, which can be leveraged to increase the system's performance over multiple \acp{CB} \cite{jabar}.
To this end, we define two sets of users in each \ac{CB}, namely \vspace{-0.05cm}
\begin{align}
 \mathcal{G}^\mathsf{P}_t &= \{k\in\mathcal{K} \,|\, \hat{\mat{H}}_{t,k}\text{ is known}\},\\[-2pt] \label{eq:unknown}
 \mathcal{G}^\mathsf{N}_t &= \{k\in\mathcal{K} \,|\, \hat{\mat{H}}_{t,k}\text{ is unknown}\},\\[-20pt]\nonumber
\end{align}
where we assume alternating CSI knowledge of $\hat{\mat{H}}_{t,k}$ between each \ac{CB}, i.e., $\mathcal{G}^\mathsf{P}_{t+1} = \mathcal{G}^\mathsf{N}_{t}$ and $\mathcal{G}^\mathsf{N}_{t+1} = \mathcal{G}^\mathsf{P}_{t}$. Moreover, we assume that the sets have the same cardinality, i.e., $|\mathcal{G}^\mathsf{P}_{t}| = |\mathcal{G}^\mathsf{P}_{t}| = \frac{K}{2} ,\, \forall t \in \{1,2\}$. Due to the known BS-IRS-user channels of one user, we are able to optimize the phase shifts at the \ac{RIS} such that they add constructively at this user. We can therefore assume that the user in $\mathcal{G}^\mathsf{P}_{t}$ in \ac{CB} $t$ has a better \textit{link strength} then the user in $\mathcal{G}^\mathsf{N}_{t}$ from a topological perspective  \cite{TSM}, if a practical number of reflect elements is deployed. Additionally, the system inherits \textit{symmetrically alternating \ac{CSI} knowledge}, as the same \ac{CB} lengths are considered. In this paper, we utilize the alternating knowledge of CSI to counteract the undesired effects of the unknown channels by employing a rate splitting strategy.

To this end, the \ac{BS} splits the requested message of each user $k$ in the group $\mathcal{G}_t^{\mathsf{P}}$ ($\mathcal{G}_t^{\mathsf{N}}$), denoted as $e_{t,k}^\mathsf{P}$ $(e_{t,k}^\mathsf{N})$ in each \ac{CB} $t$ into two sub-messages, a private part $(e_{t,k}^\mathsf{P})^{p}$ $((e_{t,k}^\mathsf{N})^{p})$ and a common part $e_{t}^c$. The \ac{BS} afterwards encodes the respective parts into a private symbol $s_{t,k}^\mathsf{P}$ ($s_{t,k}^\mathsf{N}$) and a common symbol $s_{t}^c$. After encoding the symbols, the \ac{BS} creates the beamformers of the private messages $\vect{\omega}_{t,k}^\mathsf{P}$, $\vect{\omega}_{t,k}^\mathsf{N}$ and the common message $\vect{w}_{t}^c$ and constructs the overall transmit vector $\vect{x}_t$ defined as \vspace{-0.05cm}
\begin{align}\label{eq:transVec}
   \vect{x}_t = \sum_{k\in \mathcal{G}_t^\mathsf{P}} \vect{w}_{t,k}^\mathsf{P} + \sum_{n\in \mathcal{G}_t^\mathsf{N}} \vect{w}_{t,n}^\mathsf{N} + \vect{w}_{t}^c, \\[-20pt] \nonumber
\end{align}
subject to the power constraint $\mathbb{E}\{\vect{x}_t^{\mathsf{H}}\vect{x}_t \} \leq P^\mathsf{Tr}$, where $P^\mathsf{Tr}$ is the available transmit power. Using (\ref{eq:transVec}), the power constraint can be reexpressed as \vspace{-0.1cm}
\begin{align}\label{eq:transPow}
  \sum_{k\in \mathcal{G}_t^\mathsf{P}}\hspace{-0.1cm} \norm{\vect{w}_{t,k}^\mathsf{P}}_2^2 \hspace{-0.1cm} +\hspace{-0.2cm} \sum_{n\in \mathcal{G}_t^\mathsf{N}} \hspace{-0.1cm} \norm{\vect{w}_{t,n}^\mathsf{N}}_2^2 \hspace{-0.1cm} + \hspace{-0.1cm} \norm{\vect{w}_{t}^c}_2^2 \leq  P^\mathsf{Tr},\, \forall t \in \{1,2\}.\\[-17pt]\nonumber
\end{align}
Note that $\vect{w}_{t,k}^\mathsf{P}=\zero_{L}, \forall k\in \mathcal{G}_t^\mathsf{N}$ and $\vect{w}_{t,k}^\mathsf{N}=\zero_{L}, \forall k\in \mathcal{G}_t^\mathsf{P}$, where $\zero_{L}$ denotes a column vector of length $L$ with all zero entries.

\subsection{Transmission Scheme}
We proceed to derive the ORS scheme, where, without loss of generality, we assume that the user in $\mathcal{G}_t^{\mathsf{P}}$, gets the same index assigned as the current \ac{CB}, i.e., $\mathcal{G}_1^{\mathsf{P}} = \{1\}$ and
$\mathcal{G}_2^{\mathsf{P}} = \{2\}$. At $t=1$ the transmitter sends the private message $s_{1,1}^\mathsf{P}$ intended for user 1 and $s_{1,2}^\mathsf{N}$ intended for user 2. However, the transmitter only knows the direct channels $\hat{\vect{h}}_{1,1}$ and $\hat{\vect{h}}_{1,2}$ of both users and the \ac{RIS} channels $\hat{\mat{H}}_{1,1}$ of user 1. As the \ac{RIS} channels $\hat{\mat{H}}_{1,2}$ of user 2 are unknown, as defined in (\ref{eq:unknown}), sending the private message $s_{1,1}^\mathsf{P}$ towards user 1 will cause undesirable interference at user 2. Similarly, in \ac{CB} $t=2$ the unknown reflected channels $\hat{\mat{H}}_{2,1}$ of user 1 cause undesirable interference at user 1 when transmitting the private message $s_{2,2}^\mathsf{P}$ of user 2. To tackle the problem of the undesired interference, the transmitter sends the same common message $s_{t}^c$ in both \acp{CB}, i.e., $s_{1}^c = s_{2}^c$. This enables user 1 to decode the common message in $t=1$ and use \ac{SIC} in $t=2$. Similarly, user 2 decodes the common message in $t=2$ and employs \ac{SIC} to the previously received signal in $t=1$. Using this scheme requires an \textit{opportunistic} transmission of $s_{1}^c$ in order to exploit the property of alternating \ac{CSI} knowledge as it assures that user 2 is able to decode $s_2^c$, when its channels are known, and mitigate their negative impact, when they are not.

Consequently, the user, whose channels are completely known in \ac{CB} $t$, i.e., the user in $\mathcal{G}_t^\mathsf{P}$, first decodes the common message and then utilizes the successive decoding strategy before decoding its private message. For this case, we can formulate the received signals for $k\in \mathcal{G}_t^\mathsf{P}$, $n\in \mathcal{G}_t^\mathsf{N}$ and $\forall t\in \{1,2\}$ as \vspace{-0.2cm}
\begin{align}\label{eq:recSigP}
   r_{t,k}^\mathsf{P}= & \underbrace{\big(\vect{h}_{t,k}^\mathsf{eff}\big)^\mathsf{H} \big(\vect{w}_{t,k}^\mathsf{P} +  \vect{w}_{t}^c \big)}_\text{signals that are decoded} +\underbrace{\big(\vect{h}_{t,k}^\mathsf{eff}\big)^\mathsf{H} \vect{w}_{t,n}^\mathsf{N} + v_{t,k}}_\text{interference plus noise},\\[-20pt] \nonumber
\end{align}
where $\vect{h}_{t,k}^{\mathsf{eff}} = \vect{h}_{t,k} + \mat{H}_{t,k}\vect{\theta}_t$ denotes the combination of the direct and reflected channels of user $k$ in \ac{CB} $t$ as an effective channel. In contrast, the user, whose reflected channels are unknown, employs \ac{SIC} based on the common message decoded in the other \ac{CB}, which enables the formulation of the following received signal for $k\in \mathcal{G}_t^\mathsf{P}$, $n\in \mathcal{G}_t^\mathsf{N}$ and $\forall t\in \{1,2\}$  \vspace{-0.2cm}
\begin{align}\label{eq:recSigN}
   r_{t,n}^\mathsf{N}& = \underbrace{\big(\vect{h}_{t,n}^\mathsf{eff}\big)^\mathsf{H} \big( \vect{w}_{t,n}^\mathsf{N}\big)}_\text{signals that are decoded} + \underbrace{\big(\vect{h}_{t,n}^\mathsf{eff}\big)^\mathsf{H} \vect{w}_{t,k}^\mathsf{P} + v_{t,n}}_\text{interference plus noise}.\\[-20pt] \nonumber
\end{align}
Let $\gamma_{t,k}^\mathsf{P}$ ($\gamma_{t,n}^\mathsf{N}$) denote the \ac{SINR} of the user in $\mathcal{G}_t^\mathsf{P}$ ($\mathcal{G}_t^\mathsf{N}$) decoding its private message and let $\gamma_{t,k}^c$ denote the \ac{SINR} of the user in $\mathcal{G}_t^\mathsf{P}$ decoding the common message. Based on the equations (\ref{eq:recSigP}) and (\ref{eq:recSigN}) we can write for the case $k\in\mathcal{G}_t^\mathsf{P}$ and $n\in\mathcal{G}_t^\mathsf{N}$ \vspace{-0.2cm}
\begin{align}\label{eq:gammaP}
   \gamma_{t,k}^\mathsf{P} =& \frac{|{\big(\vect{h}_{t,k}^\mathsf{eff}\big)^\mathsf{H} \vect{\omega}_{t,k}^\mathsf{P}}|^2}{|{\big(\vect{h}_{t,k}^\mathsf{eff} \big)^\mathsf{H} \vect{\omega}_{t,n}^\mathsf{N}}|^2+ \sigma_v^2},\\[-2pt] \label{eq:gammaC}
   \gamma_{t,k}^c =& \frac{|{\big(\vect{h}_{t,k}^\mathsf{eff}\big)^\mathsf{H} \vect{\omega}_{t}^{c}}|^2}{|{\big(\vect{h}_{t,k}^\mathsf{eff}\big)^\mathsf{H} \vect{\omega}_{t,k}^\mathsf{P}}|^2+|{\big(\vect{h}_{t,k}^\mathsf{eff}\big)^\mathsf{H} \vect{\omega}_{t,n}^\mathsf{N}}|^2+ \sigma_v^2}, \\ \label{eq:gammaN}
   \gamma_{t,n}^\mathsf{N} =& \frac{|{\big(\vect{h}_{t,n}^\mathsf{eff}\big)^\mathsf{H} \vect{\omega}_{t,n}^\mathsf{N}}|^2}{ |{\big(\vect{h}_{t,n}^\mathsf{eff}\big)^\mathsf{H} \vect{\omega}_{t,k}^\mathsf{P}}|^2+ \sigma_v^2}.\\[-20pt] \nonumber
\end{align}
The total achievable net rate within two \acp{CB} is then \vspace{-0.2cm}
\begin{align} \label{eq:netRate}
   R^\mathsf{net} = \frac{1}{2} \sum_{t=\{1,2\}}\overbrace{\sum_{k\in\mathcal{G}_t^\mathsf{P}} R_{t,k}^\mathsf{P} + \sum_{n\in\mathcal{G}_t^\mathsf{N}} R_{t,n}^\mathsf{N} + \frac{1}{2}R^c}^{R_t},\\[-20pt] \nonumber
\end{align}
where $R^c={\min}_{{t\in\{1,2\}},k\in\mathcal{G}_t^\mathsf{P}}\{R_{t,k}^c\}$ is shared by both users such that each user $k$ is allocated a portion $C_k$, i.e., $R_c=\sum_{k\in\mathcal{K}}{C_k}$ is satisfied. Moreover, to guarantee that the common rate $R_c$ is successfully decoded by both users in both \acp{CB}, the actual transmission rate in $t=1$ must not exceed the rate in $t=2$, i.e., $R_{1,1}^c \leq R_{2,2}^c$. Let $B$ denote the \ac{DL} transmission bandwidth, $T_\mathsf{coh}$ denote the number of symbols within one \ac{CB}, $\tau$ denote the number of symbols required for channel estimation and $B^\mathsf{DL} = B(1-\frac{\tau}{T_\mathsf{coh}})$. The rates introduced in (\ref{eq:netRate}) satisfy the following achievability conditions:
\begin{align}\label{eq:rateP}
R_{t,k}^\mathsf{P}&\leq B^\mathsf{DL} \log_2(1+\gamma_{t,k}^\mathsf{P}), \forall k \in \mathcal{G}_t^\mathsf{P}\,, \forall t \in\{1,2\},\\ \label{eq:rateC}
R_{t,k}^c&\leq  B^\mathsf{DL}\log_2(1+\gamma_{t,k}^c), \forall k \in \mathcal{G}_t^\mathsf{P}\,, \forall t \in\{1,2\},\\ \label{eq:rateN}
R_{t,n}^\mathsf{N}&\leq  B^\mathsf{DL} \log_2(1+\gamma_{t,n}^\mathsf{N}), \forall n \in \mathcal{G}_t^\mathsf{N}, \forall t \in\{1,2\}.
\end{align}

\section{Problem Formulation and Optimization Framework}
\subsection{Problem Formulation}
This paper considers maximizing the achievable net rate, while adressing important practical issues, i.e., \ac{CE} time, \ac{CE} errors and causality of the formulated problem. The key observation leading to our proposed solution is the fact that it becomes challenging to incorporate the unknown channels directly into the optimization framework. More precisely, note that for one user in each \ac{CB} the estimated effective channel misses the unknown reflected channels, specifically
\begin{align}\label{eq:estChan}
\hat{\vect{h}}_{t,j}^\mathsf{eff} = \begin{cases} \hat{\vect{h}}_{t,j} + \hat{\mat{H}}_{t,j} \vect{\theta}_t&,\,  j\in\mathcal{G}_t^\mathsf{P} \\
      \hat{\vect{h}}_{t,j}&, \, j\in\mathcal{G}_t^\mathsf{N}. \end{cases}
\end{align}
This lack of information introduces a tradeoff between the allocation of power towards $\vect{\omega}_{t,k}^\mathsf{P}$ and $\vect{\omega}_{t}^c$. On the one hand increasing the power of $\vect{\omega}_{t,k}^\mathsf{P}$ will have a negative impact on $R_{t,n}^\mathsf{N}$ as the unknown channels cause interference in (\ref{eq:gammaN}). On the other hand, although $\vect{\omega}_{t}^c$ does not cause any interference, ORS requires the same common message to be send is in both \acp{CB}, effectively halving the influence  of $R^c$ on the net rate (as seen in (\ref{eq:netRate})).
Additionally, $R^c$ is dependent on $R_{t,k}^c$ in both \acp{CB} but can only be  allocated in the first \ac{CB} in practice due to the system's causality.
Let $\vect{\omega}_t = [\{\vect{\omega}_{t,k}^\mathsf{P}\}_{k\in\mathcal{G}_t^\mathsf{P}}^\mathsf{T}, \{\vect{\omega}_{t,n}^\mathsf{P}\}_{n\in\mathcal{G}_t^\mathsf{N}}^\mathsf{T}, (\vect{\omega}_{t}^c)^\mathsf{T}]^\mathsf{T}$ denote the stacked beamformers in \ac{CB} $\overset{\phantom{{}_.}}{t}$.
Under consideration of the aspects above, we formulate the following problem for each \ac{CB} $t$: \vspace{-0.5cm}
\begin{align}\label{eq:P1}
   \underset{\vect{\theta}_t,\vect{\omega}_t}{\max} &\,  \quad \hat{R}_t  \tag{P1}\\
   \text{s.t.}  \,&\, (\ref{eq:transPow}) , (\hat{\ref{eq:rateP}})-(\hat{\ref{eq:rateN}}) \nonumber\\
   & |\theta_{t,n}| = 1, \qquad \quad \forall n \in \{1,\dots,N\},\,\forall t\in\{1,2\}, \label{eq:P1_theta}\\
   &\hat{R}_{1,k}^c - \hat{R}_{2,j}^c \leq 0 ,\,  \forall k\in\mathcal{G}_{1}^\mathsf{P},\,\forall j\in\mathcal{G}_2^\mathsf{P},\,\forall t\in\{2\},\label{eq:P1_commonRate}\\[-18pt] \nonumber
\end{align}
where $(\hat{\mathord{\cdot}})$ denotes the usage of the estimated (and missing) channels $\hat{\vect{h}}_{t,j}^\mathsf{eff}$ according to (\ref{eq:estChan}) instead of $\vect{h}_{t,j}^\mathsf{eff}$. Here, the unit-modulus constraints in (\ref{eq:P1_theta}) ensure that the \ac{RIS} only applies phase shifts to the reflected signal, while constraint (\ref{eq:P1_commonRate}) guarantees that the common rate in $t=2$ is always able to match the common rate allocated in $t=1$. Note that (\ref{eq:P1}) defines two temporally-uncoupled problems, which satisfy causality, as (\ref{eq:P1_commonRate}) only applies for $t=2$. However, problem (\ref{eq:P1}) mathematically captures the redundant nature of $R_c$, but ignores the interference the unknown channels may cause because they are unaccounted for in (\ref{eq:estChan}). Consequently, the allocation of resources towards $R_c$ becomes sub-optimal, when solving (\ref{eq:P1}) in $t=1$. To address this problem, this paper considers a predefined portion of the achievable private rate $R_{1,k}^\mathsf{P}$ in $t=1$ to be transmitted \textit{opportunistically} as common rate instead, allowing the users to mitigate part of the interference the \ac{BS} is unaware of. To this end, by denoting $\alpha^\mathsf{ORS}$ as ORS-ratio, we extend (\ref{eq:P1}) with the following constraint: \vspace{-0.15cm}
\begin{align}\label{P1_final}
   \underset{\vect{\theta}_t,\vect{\omega}_t}{\max} &\,  \quad \hat{R}_t  \tag{P1'}\\[-2pt]
   \text{s.t.}  \,&\, (\ref{eq:transPow}) , (\hat{\ref{eq:rateP}})-(\hat{\ref{eq:rateN}}), (\ref{eq:P1_theta}) ,(\ref {eq:P1_commonRate}) \nonumber\\
   &\alpha^\mathsf{ORS} \hat{R}_{t,k}^\mathsf{P} -  \hat{R}_{t,k}^c \leq 0, \forall k\in\mathcal{G}_t^\mathsf{P},\, \forall t \in \{1\}.\\[-20pt] \nonumber
\end{align}
\subsection{Optimization Framework}
To deal with the non-convexity of the rate constraints $({\ref{eq:rateP}})-({\ref{eq:rateN}})$, we rewrite them by introducing slack variables $\{{\xi}_{t,k}^c\}_{{t\in\{1,2\}}}^{k\in\mathcal{G}_t^\mathcal{P}}$, $\{{\xi}_{t,k}^\mathsf{P}\}_{{t\in\{1,2\}}}^{k\in\mathcal{G}_t^\mathcal{P}}$, $\{{\xi}_{t,n}^\mathsf{N}\}_{{t\in\{1,2\}}}^{n\in\mathcal{G}_t^\mathcal{N}}$ for the rates and  $\{{\beta}_{t,k}^c\}_{{t\in\{1,2\}}}^{k\in\mathcal{G}_t^\mathcal{P}}$, $\{{\beta}_{t,k}^\mathsf{P}\}_{{t\in\{1,2\}}}^{k\in\mathcal{G}_t^\mathcal{P}}$, $\{{\beta}_{t,n}^\mathsf{N}\}_{{t\in\{1,2\}}}^{n\in\mathcal{G}_t^\mathcal{N}}$  for the \acp{SINR} as
\begin{align}\label{eq:rateP_re}
\xi_{t,k}^\mathsf{P}\leq B^\mathsf{DL} \log_2(1+\beta_{t,k}^\mathsf{P}), \quad \forall k \in \mathcal{G}_t^\mathsf{P}, \forall t \in\{1,2\},&\\ \label{eq:rateC_re}
\xi_{t,k}^c\leq  B^\mathsf{DL}\log_2(1+\beta_{t,k}^c), \quad \forall k \in \mathcal{G}_t^\mathsf{P}, \forall t \in\{1,2\},&\\ \label{eq:rateN_re}
\xi_{t,n}^\mathsf{N}\leq  B^\mathsf{DL} \log_2(1+\beta_{t,n}^\mathsf{N}), \quad \hspace{-0.02cm} \forall n \in \mathcal{G}_t^\mathsf{N}, \forall t \in\{1,2\},&\\
\vect{\xi}_{t} \geq 0,\,
\vect{\beta}_{t}\geq 0, \qquad \qquad \forall t \in\{1,2\},&  \label{eq:slackGamma_pos}\\ \label{eq:SINR_P_re}
\beta_{t,k}^\mathsf{P} \leq {\gamma}_{t,k}^\mathsf{P}, \quad\forall k \in \mathcal{G}_t^\mathsf{P}, \forall t \in\{1,2\},&\\ \label{eq:SINR_C_re}
\beta_{t,k}^c\leq {\gamma}_{t,k}^c, \quad\forall k \in \mathcal{G}_t^\mathsf{P}, \forall t \in\{1,2\},&\\ \label{eq:SINR_N_re}
\beta_{t,n}^o\leq  {\gamma}_{t,n}^\mathsf{N}, \quad \hspace{-0.02cm}\forall n \in \mathcal{G}_t^\mathsf{N}, \forall t \in\{1,2\},&
\end{align}
where (\ref{eq:slackGamma_pos}) captures that all the introduced slack variables are non-negative, i.e., $\vect{\beta}_{t},\vect{\xi}_t$ are defined as stacked vectors of the introduced slack variables in \ac{CB} $t$.
Due to the coupling of $\vect{\omega}_t$ and $\vect{\theta}_t$ in $(\ref{eq:SINR_P_re})-(\ref{eq:SINR_N_re})$, we continue by utilizing an alternative optimization approach, thus decoupling the problem. This is achieved by fixing one of the variables, while optimizing the other.
\subsection{Beamforming Design}
For the duration of designing the beamformers, we assume the phase shifters at the \ac{RIS} to be fixed, enabling the removal of the constraint (\ref{eq:P1_theta}) due to its sole dependency on $\vect{\theta}_t$. By plugging  $({\ref{eq:gammaP}})-({\ref{eq:gammaN}})$ (which are now only dependent on $\vect{\omega}_t$) in the constraints $(\ref{eq:SINR_P_re})-(\ref{eq:SINR_N_re})$, they can be
approximated by using the first-order Taylor approximation around a feasible point $(\tilde{\vect{\omega}}_t), \forall t\in\{1,2\}$ as follows \cite[$(23)-(30)$]{synergConf}\vspace{0.05cm}
\begin{align}\label{eq:SINR_PN_Taylor}
&\frac{|(\vect{h}_{t,k}^\mathsf{eff})^\mathsf{H}\vect{\omega}_{t,k}^o|^2} {\beta_{t,k}^o} \geq \frac{2\text{Re}\{ (\tilde{\vect{\omega}}_{t,k}^o)^\mathsf{H}\vect{h}_{t,k}^\mathsf{eff} (\vect{h}_{t,k}^\mathsf{eff})^\mathsf{H} \vect{\omega}_{t,k}^o \}}{\tilde{\beta}_{t,k}^o} \, - \\[-2pt]
 &\quad\,{|(\vect{h}_{t,k}^\mathsf{eff})^\mathsf{H} \vect{\omega}_{t,k}^o|^2}\beta_{t,k}^o / {(\tilde{\beta}_{t,k}^o)^2} = f_{t,k}^o  ,\,\forall o\in\{\mathsf{P,N}\} ,\, \forall k\in\mathcal{G}_t^o,\nonumber\\[2pt] \label{eq:SINR_P_Taylor}
&\frac{|(\vect{h}_{t,k}^\mathsf{eff})^\mathsf{H}\vect{\omega}_{t}^c|^2} {\beta_{t,k}^c} \geq \frac{2\text{Re}\{ (\tilde{\vect{\omega}}_{t}^c)^\mathsf{H}\vect{h}_{t,k}^\mathsf{eff} (\vect{h}_{t,k}^\mathsf{eff})^\mathsf{H} \vect{\omega}_{t}^c \}}{\tilde{\beta}_{t,k}^c}\,- \\[-2pt]
&\quad\,{|(\vect{h}_{t,k}^\mathsf{eff})^\mathsf{H} \vect{\omega}_{t}^c|^2}\beta_{t,k}^c / {(\tilde{\beta}_{t,k}^c)^2} = f_{t,k}^c ,\, \forall k\in\mathcal{G}_t^\mathsf{P}.\nonumber
\end{align}
Thus, we write the convex approximations of $({\ref{eq:SINR_P_re}})-({\ref{eq:SINR_N_re}})$ as\vspace{0.05cm}
\begin{align}\label{eq:compTalyorP}
 {|{\big(\vect{h}_{t,k}^\mathsf{eff}\big)^\mathsf{H} \vect{\omega}_{t,n}^\mathsf{N}}|^2+ \sigma_v^2} - f_{t,k}^\mathsf{P},\,\,\forall k \in \mathcal{G}_t^\mathsf{P},&\forall n \in \mathcal{G}_t^\mathsf{N},\\ \label{eq:compTalyorN}
 {|{\big(\vect{h}_{t,n}^\mathsf{eff} \big)^\mathsf{H} \vect{\omega}_{t,k}^\mathsf{P}}|^2+ \sigma_v^2} - f_{t,n}^\mathsf{N},\,\,\forall k \in \mathcal{G}_t^\mathsf{P},&\forall n \in \mathcal{G}_t^\mathsf{N},\\ \label{eq:compTalyorC}
  {|{\big(\vect{h}_{t,k}^\mathsf{eff}\big)^\mathsf{H} \vect{\omega}_{t,k}^\mathsf{P}}|^2+|{\big(\vect{h}_{t,k}^\mathsf{eff}\big)^\mathsf{H} \vect{\omega}_{t,n}^\mathsf{N}}|^2+ \sigma_v^2} - f_{t,k}^c,& \nonumber \\ \forall k \in \mathcal{G}_t^\mathsf{P},&\forall n \in \mathcal{G}_t^\mathsf{N}.\\[-15pt] \nonumber
\end{align}
With the approximations defined above, the problems
(\ref{P1_final}) when optimizing the beamformers $\vect{w}_t$ can be approximated as
\begin{align}\label{P2}
   \underset{\vect{\omega}_t,\vect{\xi}_t,\vect{\beta}_t}{\max} &\,   \xi_t^\mathsf{obj} = {\sum}_{k\in\mathcal{G}_t^\mathsf{P}}( \xi_{t,k}^\mathsf{P} + \frac{\rho_t}{2}\xi_{t,k}^c) + {\sum}_{n\in\mathcal{G}_t^\mathsf{N}} \xi_{t,n}^\mathsf{N}  \tag{P2}\\[-3pt]
   \text{s.t.}  \,&\, (\ref{eq:transPow}) , ({\ref{eq:rateP_re}})-({\ref{eq:slackGamma_pos}}),(\hat{\ref{eq:compTalyorP}})-(\hat{\ref{eq:compTalyorC}}), \nonumber\\
   &\alpha^\mathsf{ORS} {\xi}_{t,k}^\mathsf{P} -  {\xi}_{t,k}^c \leq 0, \forall k\in\mathcal{G}_t^\mathsf{P},\, \forall t \in \{1\}, \label{eq:convexOpp}\\
   &{\xi}_{1,k}^c - {\xi}_{2,j}^c \leq 0 ,\,  \forall k\in\mathcal{G}_{1}^\mathsf{P},\,\forall j\in\mathcal{G}_2^\mathsf{P},\,\forall t\in\{2\}, \label{eq:convexCommon}
   \end{align}
where $\rho_t=1 $ if $t=1$, otherwise $\rho_t=0$.
\begin{figure*}[!ht]
   \centering
   \begin{minipage}{.485\textwidth}\vspace{-0.3cm}
   \phantom{x}\includegraphics[width=0.945\linewidth]{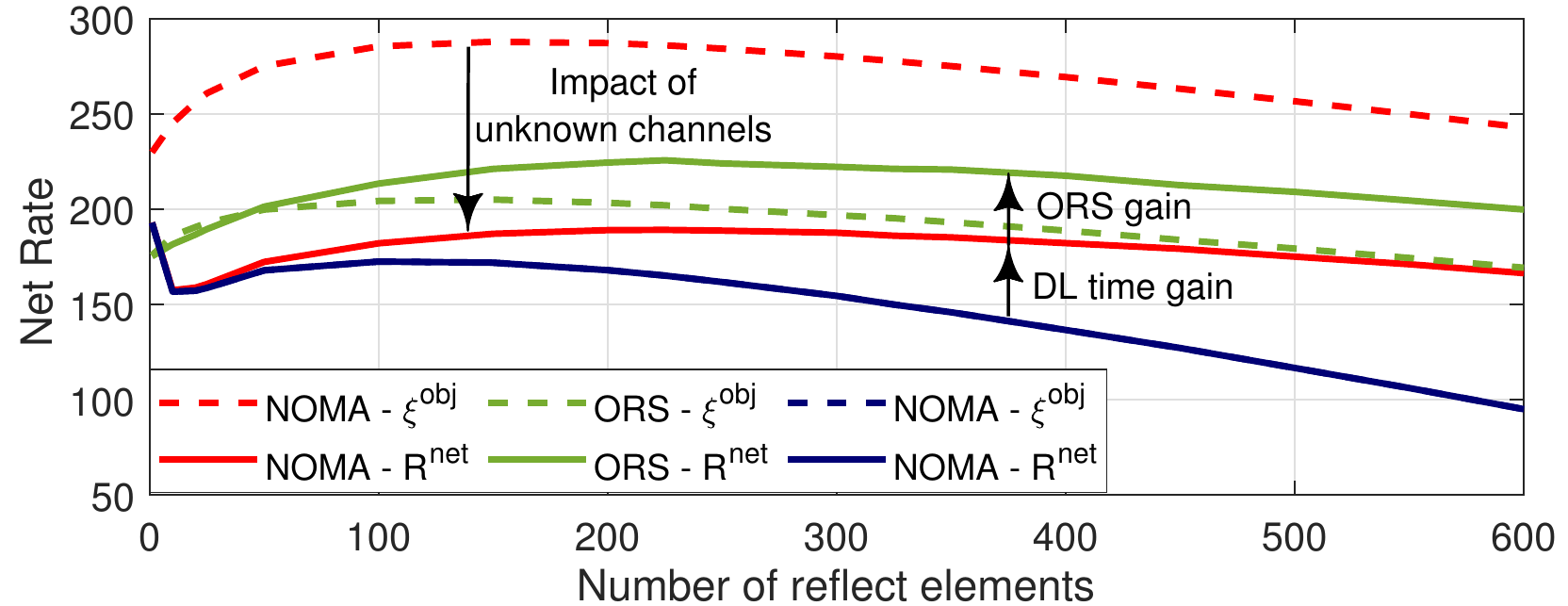}
   \begin{minipage}{.975\textwidth}\caption{Net rates for perfect CSI}\end{minipage}\begin{minipage}{.025\textwidth}\end{minipage}
   \label{fig:gains}
   \end{minipage}\hfill\begin{minipage}{.485\textwidth}
  \centering
  \includegraphics[width=0.925\linewidth]{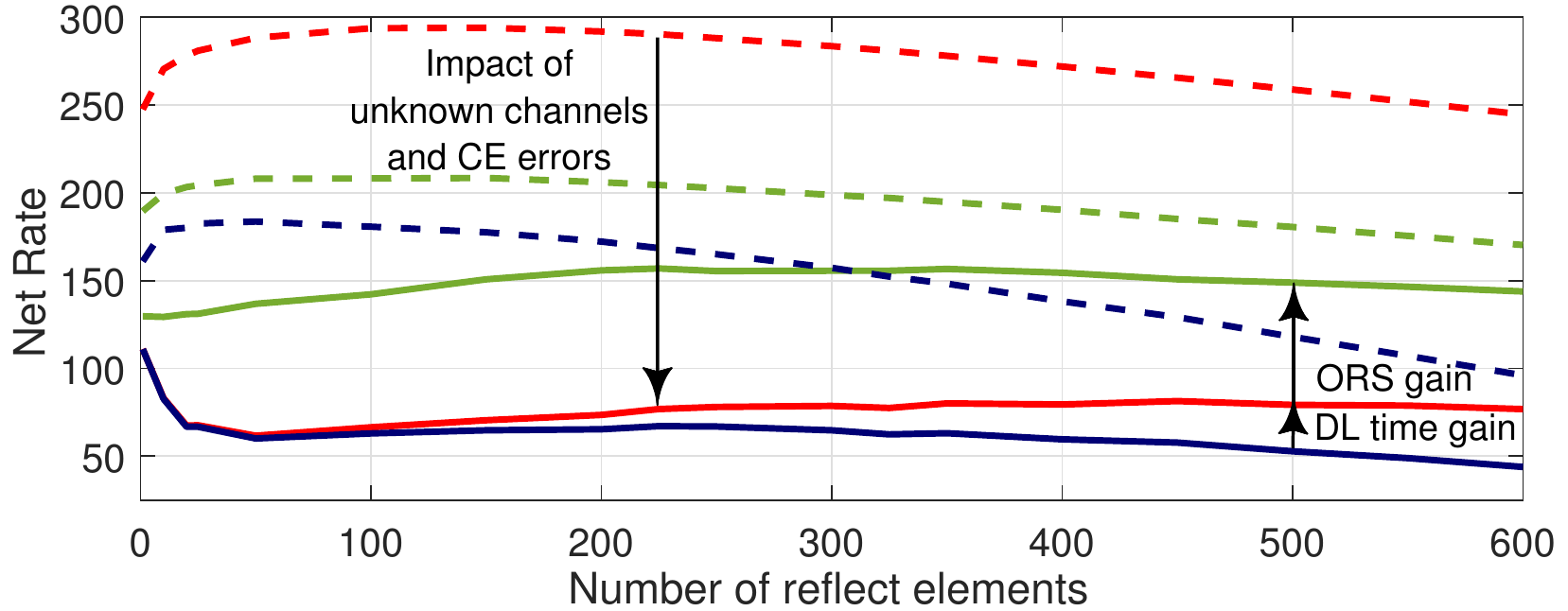}
  \begin{minipage}{0.025\textwidth}\phantom{xxx}\end{minipage}\begin{minipage}{.975\textwidth}\caption{Net rates for imperfect CSI (legend as in Fig. 1)}\end{minipage}
   \label{fig:remUsers}
   \end{minipage}
   \vspace{-0.79cm}
\end{figure*}
\subsection{Phase Shift Design}
During the design on the phase shifts we assume the beamformers $\vect{\omega}_t$ to be fixed. To obtain the optimal phase shift, we utilize the same \ac{SCA} framework, on which the beamforming design is based on. We can show that $|({\vect{h}}_{t,i}^\mathsf{eff})^\mathsf{H} \vect{\omega}_{t,k}|^2 = |\tilde{h}_{t,i,k}^o + \tilde{\mat{H}}_{t,i,k}^o\vect{\theta}_t|^2$, where $\tilde{h}_{t,i,k}^o = (\vect{\omega}_{t,k}^o)^\mathsf{H} {\vect{h}}_{t,i}$ and $\tilde{\mat{H}}_{t,i,k}^o = (\vect{\omega}_{t,k}^o)^\mathsf{H} {\mat{H}}_{t,i}$ and derive the first-order Taylor approximations for $(\ref{eq:SINR_P_re})-(\ref{eq:SINR_N_re})$ around the feasible point $(\tilde{\vect{\theta}}_t)$ in a similar fashion. Problem (\ref{P1_final}) for phase shift optimization can thus be approximated as \vspace{-0.2cm}
\begin{align}\label{P3}
  \underset{\vect{\theta}_t,\vect{\xi}_t,\vect{\beta}_t}{\max} &\,  \quad \xi_t^\mathsf{obj} - 2\kappa\big|{\sum}_{j=1}^N \theta_{t,j}^{(i-1)}(\theta_{t,j}-\theta_{t,j}^{(i-1)})\big|\tag{P3}\\[-3pt]
   \text{s.t.}  \,&\, (\ref{eq:transPow}) , ({\ref{eq:rateP_re}})-({\ref{eq:slackGamma_pos}}),(\hat{\widetilde{\ref{eq:SINR_P_re}}})-(\hat{\widetilde{\ref{eq:SINR_N_re}}}), (\ref{eq:convexOpp}),(\ref{eq:convexCommon}) \nonumber\\[-20pt] \nonumber
   \end{align}
where constraint (\ref{eq:P1_theta}) has been added as a penalty term to the objective function \cite{penalty_theta} and $\widetilde{(\cdot)}$ denotes the first order Taylor approximations around $(\tilde{\vect{\theta}}_t)$. Here, $\kappa$ is a large positive constant and the superscript $(i-1)$ denotes the value of the variable at the previous iteration.

The proposed algorithm for acquiring optimal $\vect{w}_t^*$ and $\vect{\theta}_t^*$ in each \ac{CB} by solving (\ref{P1_final}) is outlined in Algorithm 1, where we initialize the \acp{SCA} with random phase shifters and \ac{MRT} beamformers.\vspace{-0.2cm}
\begin{algorithm}
\footnotesize
\caption{\footnotesize Procedure to determine $\vect{\omega}_t^*$ and $\vect{\theta}_t^*$ of (\ref{P1_final})}\label{alg}
\begin{algorithmic}
\FOR{$t=\{1,2\}$}
   \STATE Initialize: $\kappa$, $\epsilon$, $\widehat{\vect{\theta}}_t^0$, $\widehat{\vect{\omega}}_t^0$, $i\leftarrow 0$, calculate \acp{SINR} $\vect{\beta}_t^0$ and  rates $\vect{\xi}_t^0$ \\
   \WHILE{the increase of the objective $\xi_t^\mathsf{obj}$ is above $\epsilon$}
      \STATE\hspace{-0.25cm} Obtain $\widehat{\vect{\omega}}_t^{i+1}$,$\vect{\beta}_t^{i+1}$,$\vect{\xi}_t^{i+1}$ solving (\ref{P2}) with $\vect{\theta}_t^i$,$\vect{\beta}_t^{i}$,$\vect{\xi}_t^{i}$.
      \STATE\hspace{-0.25cm} Obtain $\widehat{\vect{\theta}}_t^{i+1}$,$\vect{\beta}_t^{i+1}$,$\vect{\xi}_t^{i+1}$ solving (\ref{P3}) with $\widehat{\mat{\vect{\omega}}}_t^{i+1}$,$\vect{\beta}_t^{i+1}$,$\vect{\xi}_t^{i+1}$,$\vect{\theta}_t^i$.
   \STATE\hspace{-0.25cm} Set $i \leftarrow i+1$.
   \ENDWHILE, $\vectw_t^* \leftarrow \widehat{\vect{\omega}}_t^{i}$, $\vect{\theta}_t^* \leftarrow \widehat{\vect{\theta}}_t^{i}$
   %\STATE$\vectw_t^* \leftarrow \widehat{\vect{\omega}}_t^{i}$, $\vect{\theta}_t^* \leftarrow \widehat{\vect{\theta}}_t^{i}$
\ENDFOR
\end{algorithmic}
\end{algorithm} \vspace{-0.4cm} 
\section{Numerical results}
For the simulations we assume the \ac{BS} is equipped with $L=8$ antennas. The \ac{BS} and \ac{RIS} are assumed to be 400 m apart and facing each other, while the users are randomly positioned in a circle with a 50 m radius, whose centerpoint is 100 m away from the \ac{RIS} position. Moreover, we assume the reflect elements at the \ac{RIS} to be deployed in a rectangular grid with $\frac{\lambda}{8}$ spacing \cite{tsilipakos2020toward}, where $\lambda=0.1$ m is the wavelength. Consequently, we employ the correlated channel model introduced in \cite{corrBj}, where the average attenuation intensity is modeled after \cite[Eq.(23)]{BjonAtten}. We assume the reflected (direct) channels to be in line-of-sight (non-line-of-sight) and \ac{BS} antennas with $\frac{\lambda}{2}$ spacing.
%\begin{figure*}[!h]
%   \centering
%   \includegraphics[width=0.4\linewidth]{pictures/perf_2k-v3}
%   \caption{Perfect CSI}
%%   \end{figure}
%%   \begin{figure}[!ht]
%   \centering
%   \includegraphics[width=0.4\linewidth]{pictures/est_2k-v3}
%   \caption{Imperfect CSI}
%\end{figure*}
For the channel estimation, we consider two scenarios: 1) perfect \ac{CSI} and 2) imperfect \ac{CSI}, where we estimate the channels according to the DFT-based method in \cite{Hanzo} with $\sigma_z^2= -100$ dBm,
$P_\mathsf{UL}=30$ dBm. Further, we assume a bandwidth of $B=10$ MHz, $P^\mathsf{Tr} = 40$ dBm, $T_\mathsf{coh}=2000$ and $\sigma_v^2= -100$ dBm, $\kappa=10^4$, $\epsilon=1$. As a baseline approach, we consider \ac{NOMA}, where only the private messages are transmitted. In order to visualize the impact of the unknown channels, we compare the actual net rate $R^\mathsf{net}$ (calculated with $\vect{h}_{t,k}^\mathsf{eff}$) with the average objective function of (\ref{P1_final}) over both \acs{CB} $\xi^\mathsf{obj} = ({\hat{R}_1 +\hat{R}_2})/{2}$ (calculated with $\hat{\vect{h}}_{t,k}^\mathsf{eff}$ in (\ref{eq:estChan})). For determining $\alpha^\mathsf{ORS}$, we assume \ac{LSF} knowledge of the unknown user from the previous \ac{CB} in $t=1$, denoted by $LSF_k$, calculate $\Gamma=\max\{\frac{LSF_1}{LSF_2},\frac{LSF_2}{LSF_1}\}$ and set $\alpha^\mathsf{ORS}= (\log_2(\frac{1+\Gamma}{1+\Gamma^{-1}}))^{-1}$.

Fig. 1 compares the net rates of using ORS and NOMA with perfect \ac{CSI}. The figure shows that for a large number of reflect elements, the impact of estimating only half the channels, i.e., reducing $\tau_\mathsf{all}$ to $\tau_\mathsf{half}$ to obtain more \ac{DL} time, is beneficial even for the baseline scheme of \ac{NOMA}. When utilizing the ORS scheme, a 20\% gain over \ac{NOMA} can be observed, when a practical number for $N$ is chosen, specifically $N\geq25$. Fig. 2 depicts the impact of imperfect \ac{CSI} on the studied schemes. When compared to the curves for $R^\mathsf{net}$ in Fig. 1, it becomes apparent that the additional \ac{CE} error has a major impact on the \ac{NOMA} schemes. The rationale behind this observation is that the optimization of (\ref{P1_final}) will recover a combination of $\vect{\omega}_t$ and $\vect{\theta}_t$, which suppress interference at each user. In Fig. 1 the interference impacting $R^\mathsf{net}$ is only caused by the unknown channels. In Fig. 2 however both, the unknown channels and the estimation errors cause interference at the users, resulting in heavily reduced performance. By contrast, the negative impact on the performance of the ORS scheme is less pronounced due to the inherent robustness of rate splitting against interference. This results in gains of up to 115\%  and up to 160\% in terms of net rate for imperfect \ac{CSI}, compared to \ac{NOMA} with partial and full \ac{CSI} knowledge, respectively.
\vspace{-0.1cm}

%\begin{minipage}{0.5\linewidth}
%\begin{figure}
%%\includegraphics[width=0.7\linewidth]{pictures/perf_2k}
%\includegraphics[width=\linewidth]{pictures/est_2k-v2}
%\caption{Net rates for imperfect CSI}
%\end{figure}
%\end{minipage}\begin{minipage}{0.5\linewidth}
%\begin{figure}
%\includegraphics[width=\linewidth]{pictures/est_2k-v2}
%\caption{Net rates for imperfect CSI}
%\end{figure}
%\end{figure} 
\section{Conclusion}\label{ch:conc}
\vspace{-0.05cm}
In this paper, we propose the ORS scheme, which counteracts practical limitations of \ac{RIS} by combining opportunistic communications with \ac{RS}. By sacrificing CSI knowledge of the \ac{RIS} channels alternatively between each coherence block, we not only increase the available time for DL transmissions, but also guarantee that each user successfully mitigates the interference caused by the unknown CSI. Simulation results show that ORS provides a substantial performance uplift under practical assumptions, culminating in gains of up to 160\% over NOMA in terms of net rate. 
\balance
%\newpage
%\pagestyle{scrplain}
%\appendix
%
%\input{content/appendix}

%\cleardoublepage

%\cleardoublepage

%%

%\addcontentsline{toc}{chapter}{List of Figures}
%\listoffigures
%
%\listofalgorithms
%\cleardoublepage

%\addcontentsline{toc}{chapter}{List of Tables}
%\listoftables
%\cleardoublepage

%%
%\cleardoublepage

%\renewcommand*{\lstlistlistingname}{List of Listings}
%\lstlistoflistings
%\cleardoublepage

%\addcontentsline{toc}{chapter}{List of Formulas}
%\listofmyequations
%\cleardoublepage

%%

%\flushbottom
%\nocite*{}
% modified: /usr/share/texmf-texlive/bibtex/bst/dinat/dinat.bst

\footnotesize
\bibliographystyle{IEEEtran}
\bibliography{references}
\balance
%\section{Acronyms}
\begin{acronym}
\setlength{\itemsep}{0.1em}
\acro{AF}{amplify-and-forward}
\acro{AWGN}{additive white Gaussian noise}
\acro{B5G}{Beyond 5G}
\acro{BS}{base station}
\acro{CB}{coherence block}
\acro{CE}{channel estimation}
\acro{C-RAN}{Cloud Radio Access Network}
\acro{CMD}{common message decoding}
\acro{CP}{central processor}
\acro{CSI}{channel state information}
\acro{CRLB}{Cramér-Rao lower bound}
\acro{D2D}{device-to-device}
\acro{DC}{difference-of-convex}
\acro{DFT}{discrete Fourier transformation}
\acro{DL}{downlink}
\acro{GDoF}{generalized degrees of freedom}
\acro{IC}{interference channel}
\acro{i.i.d.}{independent and identically distributed}
\acro{IRS}{intelligent reflecting surface}
\acro{IoT}{Internet of Things}
\acro{LoS}{line-of-sight}
\acro{LSF}{large scale fading}
\acro{M2M}{Machine to Machine}
\acro{MISO}{multiple-input and single-output}
\acro{MIMO}{multiple-input and multiple-output}
\acro{MRT}{maximum ratio transmission}
\acro{MRC}{maximum ratio combining}
\acro{MSE}{mean square error}
\acro{NOMA}{non-orthogonal multiple access}
\acro{NLoS}{non-line-of-sight}
\acro{PSD}{positive semidefinite}
\acro{QCQP}{quadratically constrained quadratic programming}
\acro{QoS}{quality-of-service}
\acro{RF}{radio frequency}
\acro{RC}{reflect coefficient}
\acro{RIS}{reconfigurable intelligent surface}
\acro{RS-CMD}{rate splitting and common message decoding}
\acro{RSMA}{rate-splitting multiple access}
\acro{RS}{rate splitting}
\acro{SCA}{successive convex approximation}
\acro{SDP}{semidefinite programming}
\acro{SDR}{semidefinite relaxation}
\acro{SIC}{successive interference cancellation}
\acro{SINR}{signal-to-interference-plus-noise ratio}
\acro{SOCP}{second-order cone program}
\acro{SVD}{singular value decomposition }
\acro{TIN}{treating interference as noise}
\acro{TDD}{time-division duplexing}
\acro{TSM}{topological signal management}
\acro{UHDV}{Ultra High Definition Video}
\acro{UL}{uplink}
%\acro{M2M}{Machine to Machine}
%\acro{B5G}{Beyond 5G}
%\acro{CP}{Central Processor}
%\acro{IRS}{Intelligent Reflecting Surface}
%\acro{IoT}{Internet of Things}
%\acro{BS}{base station}
%\acro{C-RAN}{Cloud Radio Access Network}
%\acro{TIN}{treating interference as noise}
%\acro{RS}{rate splitting}
%\acro{CMD}{common message decoding}
%\acro{RS-CMD}{rate splitting and common message decoding}
%\acro{UHDV}{Ultra High Definition Video}
%\acro{LoS}{line-of-sight}
%\acro{NLoS}{non-line-of-sight}
%\acro{AF}{amplify-and-forward}
%\acro{RF}{radio frequency}
%\acro{QoS}{quality-of-service}
%\acro{QCQP}{quadratically constrained quadratic programming}
%\acro{DC}{difference-of-convex}
%\acro{IC}{interference channel}
%\acro{SIC}{Successive Interference Cancellation}
%\acro{AWGN}{Additive White Gaussian Noise}
%\acro{SINR}{signal-to-interference-plus-noise
%ratio}
%\acro{SINRs}{signal-to-interference-plus-noise
%ratios}
%\acro{MRC}{maximum ratio combining}
%\acro{D2D}{device-to-device}
%\acro{MIMO}{multiple-input and multiple-output}
%\acro{i.i.d.}{independent and identically distributed}
%\acro{SOCP}{second-order cone program}
%\acro{SDR}{semidefinite relaxation}
%\acro{SDP}{semidefinite programming}
%\acro{PSD}{positive semidefinite}
%\acro{SVD}{singular value decomposition}
\end{acronym}

\balance
\end{document}